\documentclass[aps,pra,twocolumn,superscriptaddress]{revtex4-1}
\usepackage{graphicx}
\usepackage{bm}
\usepackage{amsmath}
\usepackage{amssymb}
\usepackage{euscript}
\usepackage{verbatim}
\usepackage{fancyhdr}
\usepackage{wrapfig}
\usepackage{setspace}
\usepackage{xcolor}
\usepackage{amsfonts}
\usepackage{subfigure}
\usepackage{lipsum}
\usepackage{natbib}

\usepackage[
colorlinks=true,
urlcolor=black,
linkcolor=blue,
citecolor=blue
]{hyperref}

\def \Renyi {R\'{e}nyi }
\newcommand{\Tr}{\mathrm{Tr}}
\newcommand{\Tt}{{\mathrm{T}_2}}
\newcommand{\otn}{{\otimes n}}
\newcommand{\sA}[1]{{\scriptscriptstyle{A_{#1}}}}

\newcommand{\tket}[1]{| #1 \rangle}
\newcommand{\tbra}[1]{\langle #1 |}
\newcommand{\eqdef}{\mathrel{\overset{\makebox[0pt]{\mbox{\tiny\rm{def}}}}{=}}}

\newcommand{\mytitle}{Measuring Fermionic Entanglement: Entropy, Negativity, and Spin Structure}

\begin{document}

\title{\mytitle}

\author{Eyal Cornfeld}
\affiliation{Raymond and Beverly Sackler School of Physics and Astronomy, Tel-Aviv University, 6997801 Tel Aviv, Israel}
\affiliation{Department of Condensed Matter Physics, Weizmann Institute of Science, Rehovot 7610001, Israel}

\author{Eran Sela}
\affiliation{Raymond and Beverly Sackler School of Physics and Astronomy, Tel-Aviv University, 6997801 Tel Aviv, Israel}
\affiliation{Department of Physics and Astronomy and Quantum Materials Institute, University of British Columbia, Vancouver, B.C., V6T 1Z1, Canada }

\author{Moshe Goldstein}
\affiliation{Raymond and Beverly Sackler School of Physics and Astronomy, Tel-Aviv University, 6997801 Tel Aviv, Israel}

\begin{abstract}
The recent direct experimental measurement of quantum entanglement paves the way towards a better understanding of many-body quantum systems and their correlations. Nevertheless, the experimental and theoretical advances had so far been predominantly limited to bosonic systems. Here, we study fermionic systems. Using experimental setups where multiple copies of the same state are prepared, arbitrary order R\'{e}nyi entanglement entropies and entanglement negativities can be extracted by utilizing spatially-uniform beam splitters and on-site occupation measurement. As an example, we simulate the use of our protocols for measuring the entanglement growth following a local quench. We also illustrate how our paradigm could be used for experimental quantum simulations of fermions on manifolds with nontrivial spin structures.
\end{abstract}

\maketitle

\section{Introduction and main results}
Quantum entanglement is a profound property of many-body quantum systems~\cite{Horodecki2009}. 
Entanglement recently surfaced in many areas of quantum physics, both in condensed matter~\cite{Rev2008Amico} and in high energies~\cite{calabrese2009entanglement,nishioka2009t,Srednicki1993Entropy}. Moreover, entanglement is crucial for the investigation of strongly correlated many-body systems~\cite{Rev2008Amico} and thus a cornerstone in building effective numerical techniques for strongly correlated many-body systems~\cite{White92,Vidal2003Efficient,Schollwock2005,Verstraete2008,schollwock2011density}. It is therefore important to develop experimental protocols for the detection and characterization of entanglement.
Entanglement was recently experimentally measured in bosonic cold atoms~\cite{Islam2015Measuring,Kaufman794}, photonic chips~\cite{pitsios2017photonic}, and trapped ions~\cite{linke2017measuring,brydges2018probing}. Moreover, there have also been many theoretical proposals~\cite{Horodecki2002Method,Abanin2012Measuring,hauke2016measuring} to measure entanglement in a multitude of physical systems, such as quantum dots~\cite{Banchi2016Entanglement}, optical lattices~\cite{alves2004multipartite,daley2012measuring,pichler2013thermal,Pichler2016Measurement,gray2017measuring,Elben2018Renyi,Vermersch2018Unitary},
and Gaussian states~\cite{Weedbrook2012Gaussian}. The method of Refs.~\onlinecite{Islam2015Measuring,Kaufman794} based on many-particle interference is specifically appealing for condensed matter systems since it gives entanglement between macroscopic subsystems containing many particles.

Nearly all of experimental and theoretical advancements in entanglement measurement protocols apply either only to bosonic systems, or require the application of interaction between fermions~\cite{gray2017measuring,Elben2018Renyi,Vermersch2018Unitary}. Measurement protocols for fermionic systems had so far remained elusive due to the inherent difference of their statistics.
Nevertheless, an experimental protocol of the 2\textsuperscript{nd} \Renyi entropy entanglement measure for fermionic systems, was suggested by Pichler \textit{et al}.~\cite{pichler2013thermal}. Their result, however, was derived in methods that do not directly allow generalizations to other entanglement measures.

In this paper we present measurement protocols for fermionic systems that are directly applicable using current experimental settings~\cite{Islam2015Measuring,Kaufman794,petta2005coherent,trotzky2008time,lloyd2014quantum,folling2007direct,simon2011quantum,Theis2004Tuning,Pichler2016Measurement}. Results are presented that generalize the known bosonic results for arbitrary \Renyi entropies and negativities.
We also show how they may be used to quantum simulate fermions on manifolds with spin structures.

The generic types of systems we will consider consist of identical particles, with a special focus on fermions, hopping on lattices which are themselves partitioned into two or more subregions. We remark that entanglement emerging due to quantum statistics of identical particles, either bosonic or fermionic, is receiving attention and raising a number of fundamental issues~\cite{ghirardi2002entanglement,PhysRevLett.91.097902,Tichy_2012,PhysRevLett.121.150501}. In view of the diverse literature on this fundamental topic, our focus in this paper is treatment of the specific entanglement-measuring protocols~\cite{alves2004multipartite,daley2012measuring,pichler2013thermal,Pichler2016Measurement,gray2017measuring,GoldsteinSela2018,cornfeld2018imbalance} implemented in Refs.~\onlinecite{Islam2015Measuring,Kaufman794}.

\subsection{Entanglement Entropy}

Entanglement is naturally quantified by the entanglement entropy~\cite{Horodecki2009}, which measures the information in a subsystem with no knowledge of the remaining system. The entanglement entropy was shown to be useful in probing numerous properties of many-body quantum systems~\cite{Rev2008Amico,calabrese2009entanglement,LAFLORENCIE2016,Nielsen1999Conditions,Calabrese2008ES,GoldsteinSela2018,Nielsen2001Separable}, such as quantum critical behaviour~\cite{Vidal2003Entanglement} and non-equilibrium dynamics~\cite{Bardarson2012,daley2012measuring}. For example, the entanglement entropy of scales differently for bosonic and fermionic systems~\cite{PhysRevLett.96.010404,Eisert2010Colloquium}; furthermore, many novel states of matter that cannot be defined by their symmetries, such as topological phases~\cite{Kitaev2006,Levin2006,jiang2012identifying} and spin liquids~\cite{Zhang2011,isakov2011topological}, are discernible by their entanglement entropy scaling properties~\cite{Eisert2010Colloquium}.


For a system in a pure state \(\tket{\psi}\), its density matrix (DM) is given by \(\hat{\rho}=\tket{\psi}\tbra{\psi}\). If one bi-partitions the system into \(A\cup B\), the quantum information available to an observer in region \(A\) is encoded by the reduced DM,  \(\hat{\rho}_A=\Tr_B\hat{\rho}\). One may thus use the subsystem von-Neumann entropy \(S(A)=-\Tr\hat{\rho}_A\log\hat{\rho}_A\) to quantify the entanglement between the subsystems. If \(A\) and \(B\) are unentangled, \(\tket{\psi^{AB}}=\tket{\psi^A}\tket{\psi^B}\), then \(S(A)=0\) and \(\Tr\{\hat{\rho}_A^n\}=1\) for any integer \(n\ge1\). One may therefore use other entanglement measures such as the \Renyi entanglement entropies \cite{Horodecki2009,Mintert2007Observable},
\begin{equation}\label{rendef}
S_n(A)=\frac{1}{1-n}\log\Tr\hat{\rho}_A^n,
\end{equation}
which give various entanglement bounds~\cite{Mintert2005Concurrence,Aolita2006Measuring,mintert2007entanglement} and are directly related to the von-Neumann entropy \(S(A)=\lim_{n\to1}S_n(A)\). Moreover, even if the system \({A\cup B}\) is in a mixed state, one may use these entropies to evaluate the mutual information~\cite{Srednicki1993Entropy,Rev2008Amico,Eisert2010Colloquium} \({I_n(A:B)=S_n(A)+S_n(B)-S_n(A\cup B)}\).

\subsection{Measurement Protocols}\label{sec:meas}

Following theoretical proposals in Refs.~\onlinecite{alves2004multipartite,daley2012measuring}, a measurement of the 2\textsuperscript{nd} \Renyi entropy was realized by Islam \textit{et al}.~\cite{Islam2015Measuring}. This experimental advancement was accomplished using many-copy protocols where one creates \(n\) identical copies~\cite{Ekert2002Direct,Horodecki2002Method} of a system with a DM \(\hat{\rho}^\otn=\hat{\rho}\otimes\hat{\rho}\otimes\cdots\otimes\hat{\rho}\). The key theoretical idea put forwards by  Daley \textit{et al}.~\cite{daley2012measuring}, is that, for any \emph{bosonic} system, the entropies may be directly extracted from occupancy measurements of the particle number \(N_k^A\) in copy \(k=1\dots n\) of region \(A\).

\begin{figure}[t]
	\centering
	\includegraphics[width=1.\columnwidth]{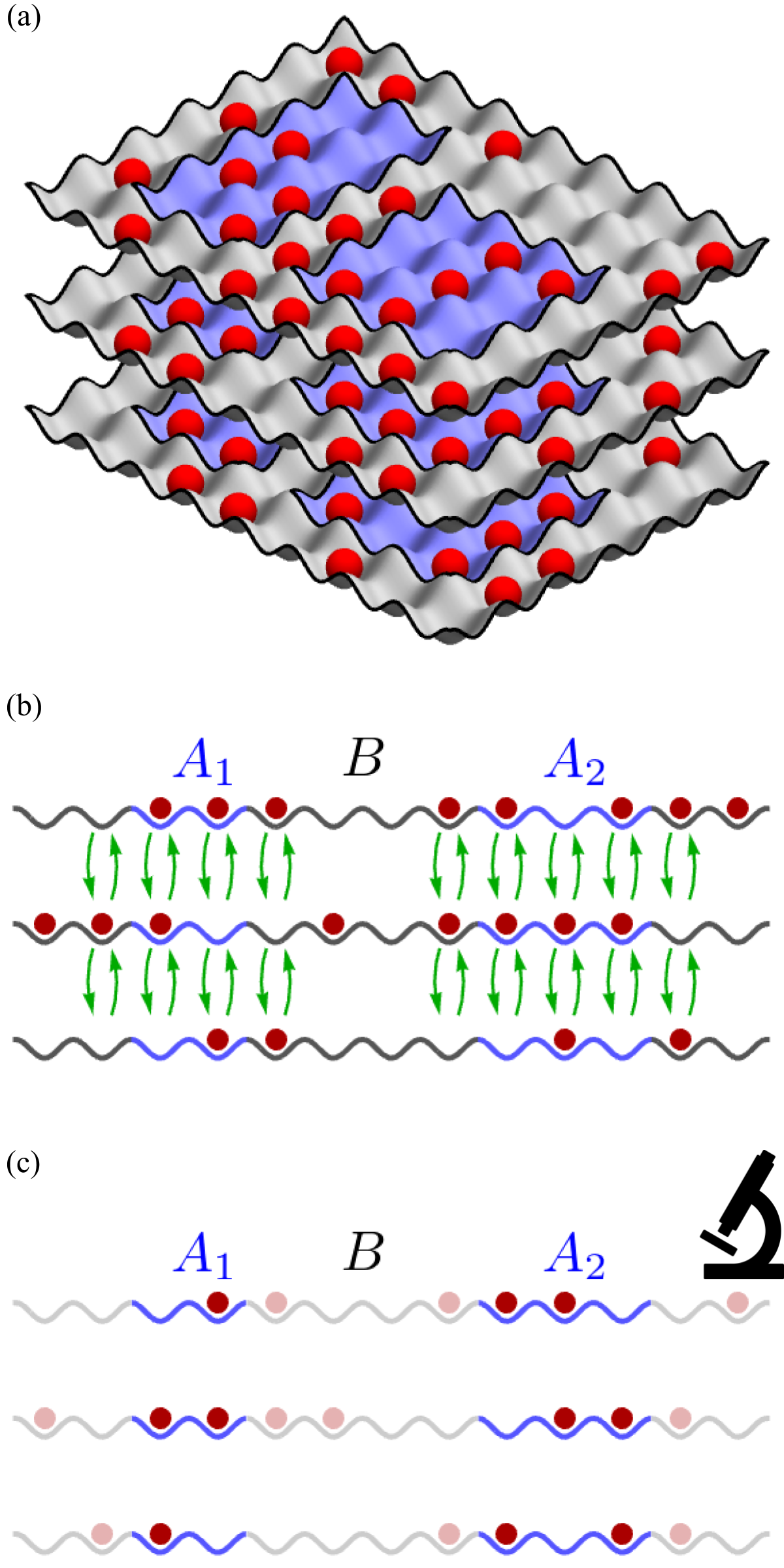}
	\caption{Schematic description of the measurement protocols, Eqs.~(\ref{mainres}), (\ref{mainres2}), for either entanglement entropy between $A=A_1\cup A_2$ and $B$, or entanglement negativity between $A_1$ and $A_2$. For discussion, see Sec.~\ref{sec:meas}.\label{fig:1}}
\end{figure}

The basic steps of the protocol are depicted in Fig~\ref{fig:1} and are as follows:

(i) Prepare \(n\) copies of a quantum system; Fig.~\ref{fig:1}(a).

(ii) Apply a unitary evolution realizing a Fourier transform (FT) in the \(n\)-copy spase; Fig.~\ref{fig:1}(b).

(iii) Measure the occupancies, \(N_k^A\), in every copy, and evaluate a function of the occupancies, \(f(\{N\})={\textstyle\prod\nolimits_{k=1}^n} e^{\frac{2\pi ik}{n}N^A_k}\); Fig.~\ref{fig:1}(c). It is this function which is affected by fermionic minus signs; see Eq.~(\ref{mainres}).

(iv) Repeat these steps and calculate the average \(\langle f(\{N\})\rangle\). 

Such an experimental protocol was carried out for \(n=2\) by Islam \textit{et al}.~\cite{Islam2015Measuring}; see Ref.~\onlinecite{daley2012measuring}. The protocol is encapsulated by the following operator relation,
\begin{equation}\label{renbos}
\Tr\hat{\rho}_A^n=\Tr\big\{f(\{\hat{N}\})\widetilde{\hat{\rho}^\otn}\big\},
\end{equation}
where, \(\widetilde{\hat{\rho}^\otn}\) is a Fourier transformed DM; see Eqs.~(\ref{FT}), (\ref{renres}) for a technical definition.

The \emph{first main result} of this paper is the novel adaptation of this protocol to \emph{fermionic} systems, where we find
\begin{equation}\label{mainres}
f(\{N\})=
\begin{cases} \delta_{N_\mathrm{avg}^A\in\mathbb{N}}~(-1)^{{N}_\mathrm{avg}^A}~\prod\limits_{k=1}^n e^{\frac{2\pi ik}{n} {N}_{k}^A} &   n~\mathrm{even},\\
\delta_{N_\mathrm{avg}^A\in\mathbb{N}}~\prod\limits_{k=1}^n e^{\frac{2\pi ik}{n} {N}_{k}^A} &   n~\mathrm{odd}.
\end{cases}
\end{equation}
Here, \({N}_\mathrm{avg}^A=\frac{1}{n}{N}_\mathrm{tot}^A=\frac{1}{n}\sum_{k=1}^n{N}_k^A\).
In fact, all protocols in this paper are to be interpreted using the same (i-iv) steps; the various entanglement measures differ only by the choice of FTs and of \( f(\{N\})\), they are also brought in a similar form to that of Eq.~(\ref{renbos}).

\subsection{Entanglement Negativity}

\begin{figure*}[t]
	\centering
	\includegraphics[width=1\linewidth]{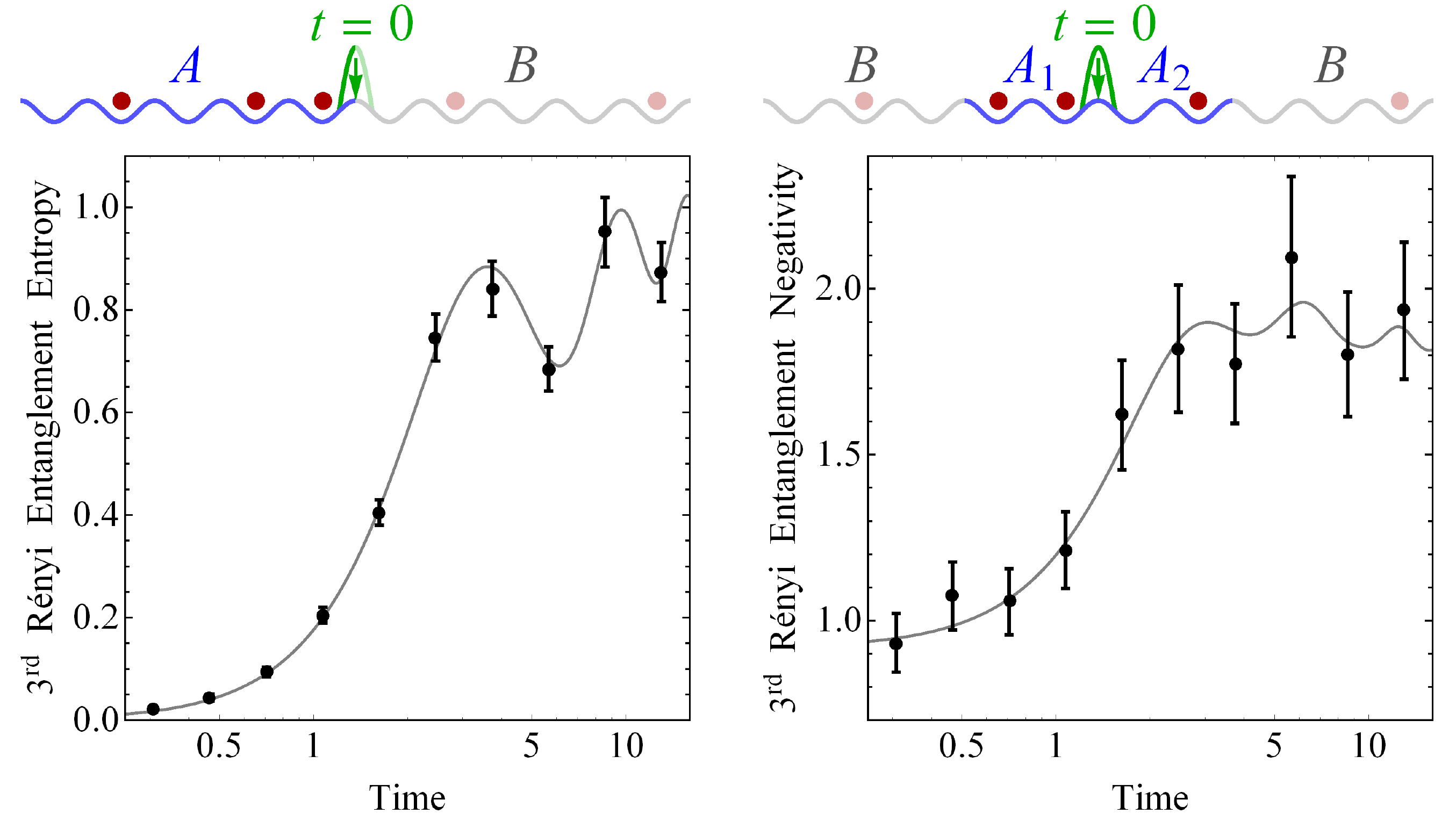}
	\caption{Simulations of our measurement protocols for the Klich-Levitov quench model; see Sec.~\ref{sec:fur}. (top) Potential barrier between two spinless leads is eliminated at ${t=0}$. (bottom) Simulated entanglement measurements at times ${t>0}$, in units of lattice spacing over Fermi velocity. Dots with error bars display protocols' outcomes, solid lines depict exact entanglements. Additional parameters are given Sec.~\ref{sec:num}.
		\label{fig:quench}}
\end{figure*}

The entropy ceases being a good measure of entanglement between two subsystems when the system is either open, mixed, or multi-partitioned. Therefore, in these generic cases, other entanglement measures must be deployed~\cite{Horodecki2009,plenio2007introduction}. Entanglement negativity \cite{Peres1996Separability} is both computationally tractable and experimentally viable~\cite{Vidal2002Computable,PhysRevA.58.883,lee2000partial,huang2014computing,gray2017measuring}. It emerged as a principal entanglement measure~\cite{Horodecki2009,plenio2007introduction,eisler2014entanglement,lanyon2017efficient,calabrese2012negativity,Rev2008Amico,LAFLORENCIE2016,Ruggiero2016Negativity,PhysRevA.60.3496,lee2000partial,eisert1999comparison,eisert2006entanglement} and was explored in various contexts ranging from numerical techniques~\cite{Ruggiero2016Negativity,Chung2014negativity,eisler2015partial,Eisler2016Entanglement} to field-theory calculations~\cite{calabrese2012negativity,calabrese2013entanglement,calabrese2014finite,coser2016towards,hoogeveen2015entanglement}.

In such generic cases, it is desirable to quantify the entanglement between two subsystems \(A=A_1\cup A_2\) coupled to an environment \(B\). Using the Peres-Horodecki entanglement criterion~\cite{Peres1996Separability}, one sees that entanglement is entailed by nonvanishing logarithmic negativity \(\mathcal{E}({A_1:A_2})=\log|\hat{\rho}_A^\Tt|\).
Here, the partial transpose \(\hat{\rho}_A^\Tt\) of a DM \(\hat{\rho}_A\) is given by transposing the indices at region $A_2$, \textit{i.e.} \(\tbra{I^{\sA{1}};J^{\sA{2}}}\hat{\rho}_A^\Tt\tket{K^{\sA{1}};L^{\sA{2}}}\!=\!\tbra{I^{\sA{1}};L^{\sA{2}}}\hat{\rho}_A\tket{K^{\sA{1}};J^{\sA{2}}}\).
In an analogous fashion to the entropic case, one may similarly study the \Renyi entanglement negativities
\begin{equation}\label{negdef}
\mathcal{E}_n({A_1:A_2})=\log\Tr\{(\hat{\rho}_A^\Tt)^n\},
\end{equation}
such that \(\mathcal{E}({A_1:A_2})=\lim_{n\to1/2}\mathcal{E}_{2n}({A_1:A_2})\).

Recently, a practical proposal, for accurately estimating the negativity in a bosonic setting, using an efficient number of measurements, 
was suggested by Gray \textit{et al}.~\cite{gray2017measuring}, utilizing a similar type of many-copy scheme~\cite{cornfeld2018imbalance}.
Our \emph{second main result} is a protocol for measuring the \Renyi negativities in \emph{fermionic} systems, using the protocol steps (i)-(iv) described above, with suitably modified FT and $f(\{N\})$; see Eqs.~(\ref{FT2}), (\ref{mainres2}).

\subsection{Further Applications and Examples}\label{sec:fur}

As the main results of this paper we obtain measurement protocols that generalize the bosonic protocols for all \Renyi entanglement entropies \(S_n(A)\thicksim\Tr\hat{\rho}_A^n\) and all \Renyi entanglement negativities \(\mathcal{E}_n({A_1:A_2})\thicksim\Tr\{(\hat{\rho}_A^\Tt)^n\}\). This is done in a manner suiting \emph{fermionic} systems, and even, as our \emph{third main result}, allows the simulation of fermions on manifolds with nontrivial spin structure; see Sec.~\ref{sec:spin}.

As an example simulating an experimental implementation of our protocols we consider the Klich-Levitov quench model~\cite{Klich2009Quantum,klich2009many,klich2008scaling,PhysRevB.85.035409,PhysRevB.83.161408,PhysRevB.80.235412,PhysRevB.91.125406,calabrese2009entanglement}, connecting two decoupled noninteracting fermionic tight-binding chains at a certain time, and tracking the resulting growth of entanglement between two subsystems thereof. 
While in this model we can easily compute the \Renyi entropies or negativities exactly, showing entanglement growth and finite size oscillations, extracting quantum averages in a real experiment with a finite sampling of the quantum measurements may become demanding. In order to account for this hurdle and demonstrate that reasonable results can be obtained with a finite sampling, we explicitly simulated a probabilistic measurements of $N^{\sA{1,2}}_k$ within the exact state. 
This is done using the techniques of Refs.~\onlinecite{Klich2002FCS,Klich2014FCS,eisler2015partial,Eisler2016Entanglement,GoldsteinSela2018,cornfeld2018imbalance} and the results are presented in Fig.~\ref{fig:quench}. 
A more detailed discussion of the model and analyses is given in Sec.~\ref{sec:num}

To simplify the discussion below we concentrate on states where the total particle number in the \emph{entire} system \(A\cup B\) is conserved. However, a closer look shows that only the conservation of total fermion parity is needed, making our results applicable to, \textit{e.g.}, mean-field superconductors~\cite{cornfeld2018entanglement}. 
Furthermore, following recent progress in resolving both entropy and negativity into symmetry sectors~\cite{GoldsteinSela2018,cornfeld2018imbalance}, our protocols can be straightforwardly generalized as to directly measure both the symmetry resolved entropy, \(S(N)\thicksim\Tr\{\delta_{\hat{N}^A,N}\hat{\rho}_A^n\}\), and negativity, \(\mathcal{E}(\Delta N)\thicksim\Tr\{\delta_{(\hat{N}^{A_1}-\hat{N}^{A_2}),\Delta N}(\hat{\rho}_A^\Tt)^n\}\).

Moreover, knowledge of the \Renyi{} entropies and negativities yields useful information about the entanglement spectrum~\cite{Horodecki2009,Nielsen1999Conditions,Calabrese2008ES,Pichler2016Measurement,GoldsteinSela2018,Nielsen2001Separable} of \(\hat{\rho}_A\) and the negativity spectrum~\cite{eisler2014entanglement,Ruggiero2016Negativity,gray2017measuring,coser2016towards} of \(\hat{\rho}_A^\Tt\). These spectra fully characterize all entanglement attributes, and even partial knowledge of the \Renyi{} entropies and negativities may be utilized to extract valuable entanglement spectra properties.

The paper is organized as follows. In Sec.~\ref{sec:entropies} we introduce the shift operators in $n$-copy spaces as our main tool to compute the \Renyi entropy. We show how fermionic signs enter this quantity, and how one can use charge conservation to correctly account for these signs in the entanglement measurement protocol. In Sec.~\ref{sec:neg} we move to the case of entanglement negativity and show that it is intrinsically more complicated. We resolve this case by utilizing the local Majorana operator formalism, and show how the fermionic signs can be accounted for by the Fourier transform. In Sec.~\ref{sec:num} we provide an example where we estimate the utility of our protocol in an experiment; we finally conclude in Sec.~\ref{sec:conc}.

\section{Entanglement Entropies}\label{sec:entropies}
\subsection{Shift Operators in $n$-copy Space and Fermionic Signs}
We begin our discussion by introducing the key player in this paper, which is the shift operator \(\hat{V}_A\). It acts on the states \(\tket{ \psi^A_{1},\psi^A_{2},\dots,\psi^A_{n}}\) in the $n$-copy Hilbert space of region $A$, by shifting them among the copies~\cite{Ekert2002Direct}, 
\begin{eqnarray}\label{Vact}
\hat{V}_A\tket{\psi_{1}^A,\psi_{2}^A,\dots,\psi_{n}^A} &=& \tket{\psi_{n}^A,\psi_{1}^A,\dots,\psi_{n-1}^A},\\
\Tr\hat{\rho}_{A}^n  &=&  \Tr\{\hat{V}_A\hat{\rho}^\otn\}.
\end{eqnarray}
One should thus come up with a protocol that measures \(\hat{V}_A\) on the \(n\)-copy system \(\hat{\rho}^\otn\) to obtain \(\Tr\hat{\rho}_A^n\) and thus the entropies, Eq.~(\ref{rendef}).
Note, that all the following derivations hold more generally for the trace of a product of different DMs, \(\Tr\{\hat{\rho}_{1A}\cdots \hat{\rho}_{nA}\}  =  \Tr\{\hat{V}_A(\hat{\rho}_1\otimes\cdots\otimes \hat{\rho}_n)\}\), and that we restrict our attention to identical \(\hat{\rho}_k\) for simplicity.

Consider a fermionic state in the occupation basis,
\begin{equation}\label{Mma}
\tket{\mathbf{M}}=\tket{\mathbf{m}_{1},\mathbf{m}_{2},\dots,\mathbf{m}_{n}} = (\hat{\mathbf{a}}^{\dagger}_1)^{\mathbf{m}_{1}}(\hat{\mathbf{a}}^{\dagger}_2)^{\mathbf{m}_{2}}\cdots(\hat{\mathbf{a}}^{\dagger}_n)^{\mathbf{m}_{n}}\tket{0}.
\end{equation}
Here, \([m_{k}]_j\in\{0,1\}\) is the occupation at site \(j\in A\) of copy \(k=1\dots n\), and \((\hat{\mathbf{a}}^{\dagger}_k)^{\mathbf{m}_{k}}\eqdef \prod_{j\in A}[\hat{a}^{\dagger}_k]_j^{[m_{k}]_j}\), where the product is taken at some fixed order. Let us investigate what becomes of the state under the action of a unitary evolution manifesting a FT in the \(k=1\dots n\) copy space,
\begin{equation}
\label{FT}
\hat{F}\hat{\mathbf{a}}^{\dagger}_{k'}\hat{F}^\dagger=\frac{1}{\sqrt{n}}\sum_{k=1}^n \omega^{k'k}\hat{\mathbf{a}}^{\dagger}_k,
\end{equation}
where \(\omega=e^{\frac{2\pi i}{n}}\) is a primitive root of unity. This transformation may be performed by evolving the system under noninteracting Hamiltonians, and its implementation may be carried out in numerous ways~\cite{Reck1994Experimental,Bovino2005Direct,folling2007direct,daley2012measuring,Pichler2016Measurement} such as a series of beam-splitters.
The evolution acts on the general state as
\begin{equation}\label{FTact}
\hat{F}\tket{\mathbf{m}_{1}, \dots, \mathbf{m}_{n}}= \Big( {\textstyle\sum\limits_{k}}\tfrac{\omega^{1k}}{\sqrt{n}}\hat{\mathbf{a}}^{\dagger}_k \Big)^{\mathbf{m}_{1}}\!\!\!\!\!\!\cdots\Big( {\textstyle\sum\limits_{k}}\tfrac{\omega^{nk}}{\sqrt{n}}\hat{\mathbf{a}}^{\dagger}_k \Big)^{\mathbf{m}_{n}}
\tket{0}.
\end{equation}
This can be related to the shift operator \(\hat{V}_A\) by looking at the phase operator \(\hat{U}_A\), used in the bosonic variant with \(f(\{N\})=\prod\nolimits_{k=1}^n \omega^{kN^A_k}\),
\begin{equation}
\label{Udef}
\hat{U}_A\eqdef{\prod_{k=1}^n}\omega^{k \hat{N}_{k}^A},\qquad \hat{U}_A\hat{\mathbf{a}}^{\dagger}_k=\omega^{k}\hat{\mathbf{a}}^{\dagger}_k\hat{U}_A.
\end{equation}
When acting on the FTed state it yields
\begin{align}
\label{steps}
& \hat{U}_A\hat{F}\tket{\mathbf{M}}
= \Big({\textstyle\sum\limits_{k}}\tfrac{\omega^{(1+1)k}}{\sqrt{n}}\hat{\mathbf{a}}^{\dagger}_{k} \Big)^{\mathbf{m}_{1}}\!\!\!\!\!\!\cdots\Big( {\textstyle\sum\limits_{k}}\tfrac{\omega^{(n+1)k}}{\sqrt{n}}\hat{\mathbf{a}}^{\dagger}_{k} \Big)^{\mathbf{m}_{n}}
\tket{0}\nonumber\\ 
& =\Big({\textstyle\sum\limits_{k}}\tfrac{\omega^{2k}}{\sqrt{n}}\hat{\mathbf{a}}^{\dagger}_{k} \Big)^{\mathbf{m}_{1}}\!\!\!\!\!\!\cdots\Big( {\textstyle\sum\limits_{k}}\tfrac{\omega^{nk}}{\sqrt{n}}\hat{\mathbf{a}}^{\dagger}_{k} \Big)^{\mathbf{m}_{n-1}}\Big({\textstyle\sum\limits_{k}}\tfrac{\omega^{1k}}{\sqrt{n}}\hat{\mathbf{a}}^{\dagger}_{k} \Big)^{\mathbf{m}_{n}}
\tket{0} \nonumber\\
& =(-1)^{|\mathbf{m}_n|(|\mathbf{m}_1|+\ldots+|\mathbf{m}_{n-1}|)}\hat{F}\tket{\mathbf{m}_{n},\mathbf{m}_{1}, \dots, \mathbf{m}_{n-1}} \nonumber\\
& =\hat{F}\hat{V}_A(-1)^{\hat{N}_n^A(\hat{N}_{\mathrm{tot}}^A-\hat{N}_n^A)}\tket{\mathbf{M}},
\end{align}
where ${|\mathbf{m}_k|=\sum_{j\in A} [m_k]_j}$. 
In exchanging the order of creation operators, we used the fermionic commutation relations and accumulated a phase in order to get the form of Eqs.~(\ref{Vact}), (\ref{FTact}).
Since this equation holds for all states $\tket{\mathbf{M}}$, one has
\begin{equation}
\label{keyrel}
\hat{V}_A=\hat{F}^\dagger\hat{U}_A\hat{F}(-1)^{\hat{N}_n^A(\hat{N}_{\mathrm{tot}}^A-\hat{N}_n^A)}.
\end{equation}

This formal result is crucial for our further analyses. In the bosonic case, where the minus sign is absent, it provides an alternative proof of the measurement protocol by Pichler \textit{et al.}~\cite{Pichler2016Measurement} . Namely, measurements of the function $f(\{N\})$ following the FT, which is nothing but $\hat{F}^\dagger \hat{U}_A \hat{F}$, yield the shift operator in the many-copy space. The latter gives the $n$-th \Renyi entropy by definition.

Below, we will use Eq.~(\ref{keyrel}) to change the function $f(\{N\})$ in order to account for the fermionic minus sign. We remark that, a priory, this is not obvious since the number operators, $\hat{N}_k^A$, determining this sign are number operators of individual copies, before the FT. The experimental protocol, however, measures the number operators only after the FT. However, as we hereafter demonstrate, conservation laws relate the number operators in different copies and allow one to measure the fermionic signs just from the total number operator.

\subsection{Conservation Laws}
Using this operator relation, we may now successfully turn our attention to entanglement measures, whereby particle conservation becomes useful~\cite{GoldsteinSela2018,cornfeld2018imbalance}. 
Let us examine the expectation value of the shift operator over the many-copy reduced density matrix,
\begin{equation}
\hat{\rho}_{A}^\otn
={\sum\limits_{\mathbf{M}',\mathbf{M}}}\tket{\mathbf{M}'}{\prod\limits_{k}}[\rho_{A}]_{\mathbf{m}'_k,\mathbf{m}^{\phantom{|}}_k}\tbra{\mathbf{M}}.
\end{equation}
A result of particle number conservation is the commutation~\cite{GoldsteinSela2018} of the density matrices with the particle number \({[\hat{\rho}_{A},\hat{N}^A]=0}\), and hence, using index notation,
\(
[\rho_{A}]_{\mathbf{m}'_k,\mathbf{m}^{\phantom{|}}_k}=[\rho_{A}]_{\mathbf{m}'_k,\mathbf{m}^{\phantom{|}}_k}\delta_{|\mathbf{m}'_k|,|\mathbf{m}^{\phantom{|}}_{k}|}
\).
By also utilizing Eq.~(\ref{Vact}) in index form,
\(
\tbra{\mathbf{M}'}\hat{V}_A\tket{\mathbf{M}}={\prod_k}\delta_{\mathbf{m}'_{k+1},\mathbf{m}^{\phantom{|}}_{k}}
\),
one thus gets
\begin{align}\label{delta1}
&\Tr\{\hat{V}_A\hat{\rho}^\otn\}
=\sum\limits_{\mathbf{M},\mathbf{M}'}\tbra{\mathbf{M'}}\hat{V}_A\tket{\mathbf{M}}\prod\limits_{k}[\rho_{A}]_{\mathbf{m}^{\phantom{|}}_k,\mathbf{m}'_k} \\ 
&=\sum\limits_{\mathbf{M},\mathbf{M}'}\tbra{\mathbf{M'}}\hat{V}_A\tket{\mathbf{M}}\prod\limits_{k}\big(\delta_{\mathbf{m}'_{k+1},\mathbf{m}^{\phantom{|}}_{k}}\delta_{|\mathbf{m}^{\phantom{|}}_k|,|\mathbf{m}'_{k}|}\big)[\rho_{A}]_{\mathbf{m}^{\phantom{|}}_k,\mathbf{m}'_k}.\nonumber
\end{align}
Upon examining these combined relations one finds
\begin{equation}
|\mathbf{m}_{1}^{\phantom{1}}|=\ldots=|\mathbf{m}_{n}^{\phantom{1}}|=|\mathbf{m}'_{1}|=\ldots=|\mathbf{m}'_{n}|,
\end{equation}
In the \(n\)-copy space, one thus find \(\hat{N}_k^A\) as good quantum numbers.  Therefore, when substituting Eq.~(\ref{keyrel}) within the trace, one may impose
\begin{equation}\label{NinN}
N_k^A\cong\frac{1}{n}N_\mathrm{tot}^A=\hat{N}_\mathrm{avg}^A\in\mathbb{N},
\end{equation}
and hence, \((-1)^{\hat{N}_n^A(\hat{N}_{\mathrm{tot}}^A-\hat{N}_1^A)}\cong(-1)^{(n-1)\hat{N}_\mathrm{avg}^A}\), \textit{i.e.},
\begin{equation}
\Tr\{\hat{V}_A\hat{\rho}^\otn\}=\Tr\{\hat{F}^\dagger\hat{U}_A\hat{F}(-1)^{(n-1)\hat{N}_\mathrm{avg}^A}\hat{\rho}^\otn\}.
\end{equation}
Moreover, since the total particle number in region $A$ is a FT invariant, \(\hat{N}_\mathrm{tot}^A=\sum_k\hat{\mathbf{a}}^{\dagger}_ {k}\cdot\hat{\mathbf{a}}_{k}^{\phantom{|}}=\hat{F}\hat{N}_\mathrm{tot}^A\hat{F}^\dagger\), we deduce
\begin{equation}\label{Ufer}
\begin{gathered}
\Tr\hat{\rho}_A^n=\Tr\{\hat{V}_A\hat{\rho}^\otn\}  =  \Tr\{\hat{U}_A^\mathrm{fer}(\hat{F}\hat{\rho}^\otn\hat{F}^\dagger)\},\\
\hat{U}_A^\mathrm{fer}\eqdef(-1)^{(n-1)\hat{N}_\mathrm{avg}^A}\prod_{k=1}^n e^{\frac{2\pi ik}{n} \hat{N}_{k}^A}.
\end{gathered}
\end{equation}

As compared to Eq.~(\ref{keyrel}), this formula which uses particle number conservation, allows the determination of fermionic signs from measurements of the total particle number. The latter is compatible with the experimental protocol for the bosonic case, namely, it does not require one to measure an additional non-commuting observable.

\subsection{Entanglement Entropy Measurement Protocols}

Using the particle number conservation constraints, 
Eq.~(\ref{Ufer}) may be recast into the following form 
to attain our \emph{first main result}, Eq.~(\ref{mainres}),
\begin{align}\label{renres}
&n~\mathrm{even:} &&\Tr\hat{\rho}_A^n=\Tr\Big\{\delta_{N_\mathrm{avg}^A\in\mathbb{N}}(-1)^{\hat{N}_\mathrm{avg}^A}\!{\textstyle\prod\limits_{k=1}^n}\! e^{\frac{2\pi ik}{n} \hat{N}_{k}^A}\widetilde{\hat{\rho}^\otn}\Big\}, \nonumber\\
&n~\mathrm{odd:} &&\Tr\hat{\rho}_A^n=\Tr\Big\{\delta_{N_\mathrm{avg}^A\in\mathbb{N}}{\textstyle\prod\limits_{k=1}^n}\! e^{\frac{2\pi ik}{n} \hat{N}_{k}^A}\widetilde{\hat{\rho}^\otn}\Big\},
\end{align}
where \(\widetilde{\hat{\rho}^\otn}=\hat{F}\hat{\rho}^\otn\hat{F}^\dagger\) is the FTed DM.

This operator identity implies that measurements of \(\hat{U}_A^\mathrm{fer}\) on the FTed system, according to the protocol of Eq.~(\ref{renbos}), yield the expectation value of $\hat{V}_A$ and hence the \Renyi entanglement entropy. 

Note that (i) by setting \(n=2\), this result reduces to the known results of Pichler \textit{et al}.~\cite{pichler2013thermal}; and (ii) for odd \(n\), the fermionic minus signs cancel out, and one recovers the known results~\cite{daley2012measuring,GoldsteinSela2018} for the bosonic case.

Let us note a relation with field theory. One may unify the even/odd expressions by absorbing the fermionic minus sign in a relabelling of the FT index $k$ as to run over half integers for even $n$, such that \(\hat{U}_A^\mathrm{fer}=\prod_{k=-(n-1)/2}^{(n-1)/2} e^{\frac{2\pi ik}{n} \hat{N}_{k}^A}\). 
This was noticed by the conformal field theory community and used to solve for the entropies in critical fermionic systems~\cite{casini2005entanglement,Cardy2008Form,casini2008analytic,casini2009entanglement,Cornfeld2017Ising}; we herein showed this to hold in fact for any (non-critical) charge conserving system.

\section{Entanglement Negativities and Spin Structures}\label{sec:neg}
Using the valuable properties of the shift operator we may now address the evaluation of the negativities, Eq.~(\ref{negdef}), of subregions \(A=A_1\cup A_2\) for fermionic systems. 
In the bosonic case, the simplest way to evaluate the negativities relies on the properties of partial transposition,
\begin{eqnarray}\label{negrel}
\Tr\{(\hat{\rho}_A^\Tt)^n\}&=&\textstyle\Tr\{\hat{V}_A(\hat{\rho}_A^\Tt)^\otn\}=\Tr\{\hat{V}_A(\hat{\rho}_A^\otn)^\Tt\} \nonumber\\
&=&\Tr\{\hat{V}_A^\Tt\hat{\rho}_A^\otn\}=\Tr\{\hat{V}_{A_1}^{\phantom{+}}\hat{V}_{A_2}^{-1}\hat{\rho}^\otn\}.
\end{eqnarray}

Similar to the analysis of \Renyi entropies in Sec.~\ref{sec:entropies}, when exploring particle number conservation constraints applying to the negativity~\cite{cornfeld2018imbalance}, one gets
\begin{align}\label{delta2}
\tbra{\mathbf{M}'^{\sA{1}}\mathbf{M}'^{\sA{2}}}\hat{V}_A^\Tt\tket{\mathbf{M}^{\sA{1}}\mathbf{M}^{\sA{2}}}&=\prod\limits_k\delta_{\mathbf{m}'^{1}_{k+1},\mathbf{m}^{1}_{k}}\delta_{\mathbf{m}'^{2}_{k-1},\mathbf{m}^{2}_{k}},\nonumber\\
\tbra{\mathbf{M}^{\sA{1}}\mathbf{M}^{\sA{2}}}\hat{\rho}_A^\otn\tket{\mathbf{M}'^{\sA{1}}\mathbf{M}'^{\sA{2}}}&\propto\prod\limits_k\delta_{|\mathbf{m}^{1}_{k}|+|\mathbf{m}^{2}_{k}|,|\mathbf{m}'^{1}_{k}|+|\mathbf{m}'^{2}_{k}|},
\end{align}
\textit{cf.}~Eq.~(\ref{delta1}). Here \({\tket{\mathbf{M}}^{\sA{\alpha}}=\tket{\mathbf{m}_{1}^\alpha,\dots,\mathbf{m}_{n}^\alpha}}\) are the occupations in the \(n\) copies of subregion \(A_{\alpha=1,2}\); see Eq.~(\ref{Mma}).
Upon examining these combined relations we find
\begin{equation}
|\mathbf{m}_n^{\sA{1}}|-|\mathbf{m}_1^{\sA{2}}|=|\mathbf{m}_1^{\sA{1}}|-|\mathbf{m}_2^{\sA{2}}|=\ldots=|\mathbf{m}_{n-1}^{\sA{1}}|-|\mathbf{m}_n^{\sA{2}}|,
\end{equation}
and so within the trace one may impose
\begin{equation}
{N_\mathrm{avg}^{A_1}-N_\mathrm{avg}^{A_2}=\frac{1}{n}(N_\mathrm{tot}^{A_1}-N_\mathrm{tot}^{A_2})}\in\mathbb{Z},
\end{equation}
\textit{cf.}~Eq.~(\ref{NinN}).
Unfortunately, this is a dead end, 
as the fermionic minus signs of Eq.~(\ref{keyrel}) are not resolved by these constraints. To proceed, we must utilize a different venue.

\subsection{Local Operators}

A useful approach towards negativities (from both the analytical and numerical perspectives) is the local operator formalism. Any operator, and hence any reduced DM, may be expanded~\cite{Vidal2003Entanglement} by a basis of local Majorana operators, \(\hat{\mathbf{a}}^{\sA{1,2}}=\tfrac{1}{2}(\hat{\bm{\gamma}}^{\sA{1,2}}+i\hat{\bm{\gamma}}'^{\sA{1,2}})\), in the form of
\begin{equation}
\hat{\rho}_A=\sum_{\bm{\mu}^{1}\bm{\mu}^{2}}w_{\bm{\mu}^{1}\bm{\mu}^{2}}(\hat{\bm{\gamma}}^{\sA{1}})^{\bm{\mu}^{1}}(\hat{\bm{\gamma}}^{\sA{2}})^{\bm{\mu}^{2}},
\end{equation}
where \([\mu^{1,2}_{(k)}]_j\in\{0,1\}\) for all Majorana operators in region \(A_{1,2}\) (of copy \(k\)). When performing such an expansion, a natural choice of transposition is~\cite{eisler2015partial}
\begin{equation}\label{T2def}
\begin{gathered}
\hat{\rho}_A^\Tt=\tfrac{1-i}{2}\hat{\rho}_A^{+}+\tfrac{1+i}{2}\hat{\rho}_A^{-},\\
\hat{\rho}_A^{\pm}\eqdef\sum_{\bm{\mu}^{1}\bm{\mu}^{2}}w_{\bm{\mu}^{1}\bm{\mu}^{2}}(\hat{\bm{\gamma}}^{\sA{1}})^{\bm{\mu}^{1}}(\hat{\bm{\gamma}}^{\sA{2}})^{\bm{\mu}^{2}}(\pm i)^{|\bm{\mu}^2|},
\end{gathered}
\end{equation}
where \(|\bm{\mu}^{1,2}|=\sum_{j\in A_{1,2}} [\mu^{1,2}]_j\).
This decomposition is highly beneficial, as originally noted in the context of Gaussian DMs~\cite{eisler2015partial,coser2015partial,coser2016towards,coser2016spin,Eisler2016Entanglement,cornfeld2018imbalance}. It was proven~\cite{eisler2015partial} that \(\hat{\rho}_A^{\pm}\) are Gaussian as well.
To appreciate this relation, let us look at the 3\textsuperscript{rd} \Renyi negativity, \(\mathcal{E}_3({A_1:A_2})\),
\begin{equation}\label{neg3}
\Tr\{(\hat{\rho}_{A}^\Tt)^3\}=-\frac{1}{2}\Tr\{(\hat{\rho}^+_A)^3\}+\frac{3}{2}\Tr\{(\hat{\rho}^+_A)^2\hat{\rho}^-_A\}.
\end{equation}
The decomposition hereby allows one to evaluate the negativity by tracing Gaussian matrices, which is a tractable task both numerically and analytically. We herein show that the decomposition is useful beyond the Gaussian case, and is valuable for any fermionic system.

Measurements of negativities \(\Tr\{(\hat{\rho}_{A}^\Tt)^n\}\) are reduced to separate evaluations of monomials, \(\Tr\{\hat{\rho}_A^{\sigma_1}\hat{\rho}_A^{\sigma_2}\cdots\hat{\rho}_A^{\sigma_n}\}\) with \(\sigma_k=\pm\), \textit{cf.} Eq.~(\ref{neg3}). Let us begin by examining the simplest case of a pure monomial \(\Tr\{(\hat{\rho}_{A}^\pm)^n\}\).

Key properties of the partial transposition, Eq.~(\ref{T2def}), are that \((\hat{N}^{A_2})^\Tt=|A_2|-\hat{N}^{A_2}\), where $|A_2|$ is the number of sites in region $A_2$, and that therefore~\cite{cornfeld2018imbalance}
\begin{equation}
0=[\hat{\rho}_A,\hat{N}^A]^\Tt=[\hat{\rho}_A^\Tt,\hat{N}^{A_1}-(\hat{N}^{A_2})^\Tt]=[\hat{\rho}_A^\Tt,\hat{N}^A].
\end{equation}
It is straightforward to check that \([\hat{\rho}_A^\pm,\hat{N}^A]=0\) by construction. We hence recover charge conservation, which was the only prerequisite for our entanglement entropy results. We thus get \(\Tr\{(\hat{\rho}_{A}^\pm)^n\}=\Tr\{\hat{U}_A^\mathrm{fer}\hat{F}(\hat{\rho}_A^\pm)^\otn\hat{F}^\dagger\}\), by applying Eq.~(\ref{Ufer}). Following a rather technical Majorana-operator calculation (see Appendix~\ref{app:pure}) one finds that within the trace \(\hat{U}_{A_2}^\mathrm{fer}\cong(\hat{U}_{A_2}^\mathrm{fer})^{-1}(-1)^{\frac{1}{2}\sum_{k=1}^n|\bm{\mu}^{2}_k|}\). This cancels the excess phases \((\pm i)\) of the decomposed DMs Eq.~(\ref{T2def}) and restores the structure of the negativity measurements, Eq.~(\ref{negrel}),
\begin{align}\label{Uneg}
&\Tr\{(\hat{\rho}_{A}^\pm)^n\}=  \Tr\{\delta_{(N_\mathrm{avg}^{A_1}-N_\mathrm{avg}^{A_2})\in\mathbb{Z}}\hat{U}_{A}^\mathrm{neg}(\hat{F}\hat{\rho}^\otn\hat{F}^\dagger)\},\nonumber\\
&\hat{U}_A^\mathrm{neg}\eqdef\hat{U}_{A_1}^\mathrm{fer}(\hat{U}_{A_2}^\mathrm{fer})^{-1}\\
&\phantom{\hat{U}_A^\mathrm{neg}}=(-1)^{(n-1)\big(\hat{N}_{\mathrm{avg}}^{A_1}-\hat{N}_{\mathrm{avg}}^{A_2}\big)}\prod_{k=1}^n e^{\frac{2\pi ik}{n} \big(\hat{N}_{k}^{A_1}-\hat{N}_{k}^{A_2}\big)}.\nonumber
\end{align}
This result incorporates the particle number conservation constraints and directly generalizes both the known bosonic negativity measurement schemes~\cite{gray2017measuring,cornfeld2018imbalance} and our fermionic entropy results, Eqs.~(\ref{mainres}), (\ref{Ufer}), (\ref{renres}). When applying the general protocol of Eq.~(\ref{renbos}), it implies that measurements of \(\hat{U}_A^\mathrm{fer}\) on the FTed system yields the pure monomial part of the negativity; \textit{e.g.}, \(\Tr\{(\hat{\rho}^+_A)^3\}\) of Eq.~(\ref{neg3}). Note, that these pure monomials are in fact equivalent to the recently proposed partial time-reversal negativity entanglement measure~\cite{Shapourian2017Partial}. This partial time-reversal appears in the topological characterization of interacting fermionic systems and constitutes a signature of many-body topological phases~\cite{Shapourian2017Many,Shiozaki2018Many}.

\subsection{Generalized Fourier Transform}

All that is left in order to complete the protocol for the measurement of the full negativities is the treatment of mixed monomials \(\Tr\{\hat{\rho}_A^{\sigma_1}\hat{\rho}_A^{\sigma_2}\cdots\hat{\rho}_A^{\sigma_n}\}\), \textit{e.g.} \(\Tr\{(\hat{\rho}^+_A)^2\hat{\rho}^-_A\}\) of Eq.~(\ref{neg3}). This may be accomplished by considering the parity operator, \(\hat{P}^{A_2}=(-1)^{\hat{N}^{A_2}}\),
which relates the decomposed matrices \(\hat{\rho}_A^\mp=\hat{P}^{A_2}\hat{\rho}_A^\pm\hat{P}^{A_2}\). By deploying the parity on appropriate copies, one finds
\begin{equation}
\Tr\{\hat{\rho}_A^{\sigma_1}\hat{\rho}_A^{\sigma_2}\cdots\hat{\rho}_A^{\sigma_n}\}  =\Tr\{\hat{V}_A\hat{P}^{A_2}_{\{\sigma\}}(\hat{\rho}_A^{+})^\otn \hat{P}^{A_2}_{\{\sigma\}}\},
\end{equation}
where 
\(\hat{P}^{A_2}_{\{\sigma\}} \eqdef\prod_{k=1}^n(\hat{P}^{A_2}_k)^{\delta_{\sigma_1,(-)}}\). 
We thus get a generalization of the pure monomial result of Eq.~(\ref{Uneg}),
\begin{equation}\label{mixed}
\Tr\{\hat{\rho}_A^{\sigma_1}\cdots\hat{\rho}_A^{\sigma_n}\!\}  \!=\!  \Tr\{\delta_{(N_\mathrm{avg}^{A_1}-N_\mathrm{avg}^{A_2})\in\mathbb{Z}}\hat{U}_{A}^\mathrm{neg}(\hat{F}_{\{\sigma\}}^{\phantom{|}}\hat{\rho}^\otn\hat{F}_{\{\sigma\}}^{\dagger})\}.
\end{equation}
Here, we utilized a generalized FT, \(\hat{F}_{\{\sigma\}}\eqdef\hat{F}^{A_{1}}_{\phantom{|}}\hat{F}^{A_{2}}_{\phantom{|}} \hat{P}^{A_{2}}_{\{\sigma\}}\), which satisfies
\begin{equation}
\label{FT2}
\hat{F}_{\{\sigma\}}^{\phantom{|}}\hat{\mathbf{a}}^{A_{2}\dagger}_{k'}\hat{F}_{\{\sigma\}}^{\dagger}=\frac{1}{\sqrt{n}}\sum_{k=1}^n \sigma_{k'}\omega^{k'k}\hat{\mathbf{a}}^{A_{2}\dagger}_k,
\end{equation}
\textit{cf.} Eqs. (\ref{FT}), (\ref{Udef}). This modified FT may be readily achieved using similar unitary evolution protocols applied for the standard FT~\cite{Reck1994Experimental,Bovino2005Direct,folling2007direct}.

\begin{figure}[t]
	\centering
	\includegraphics[width=1.\columnwidth]{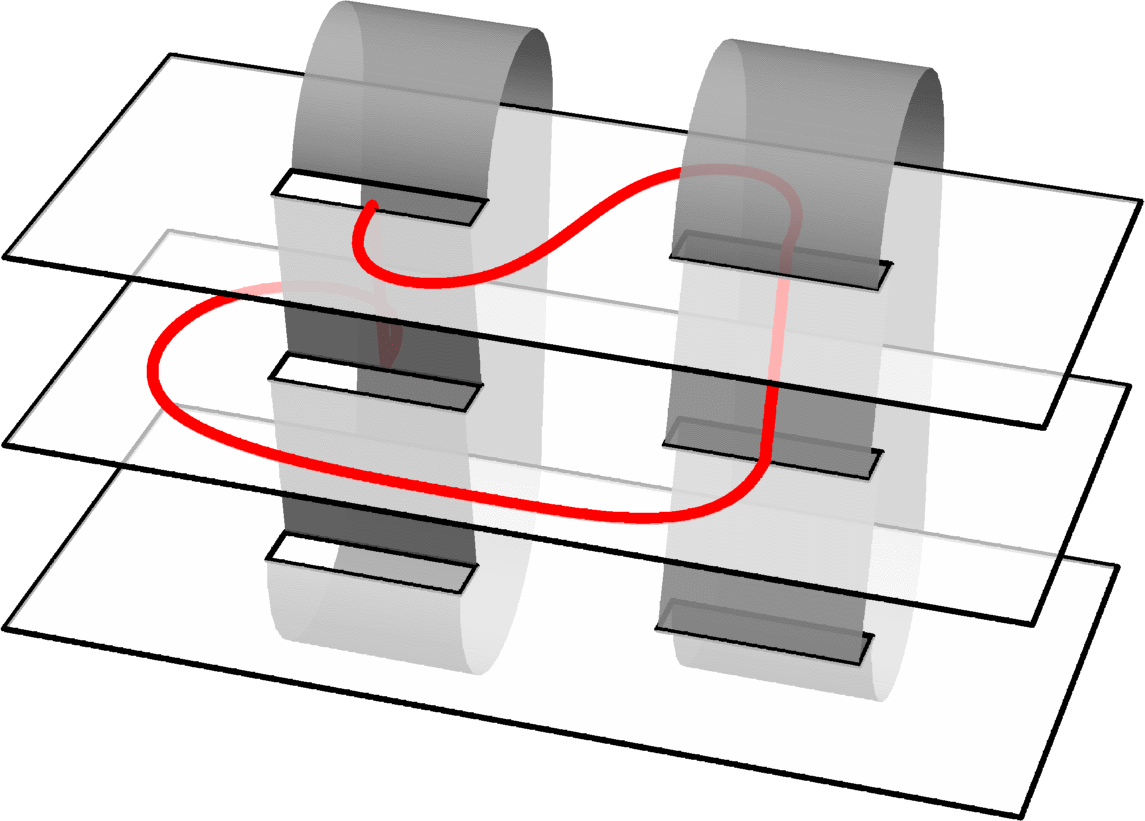}
	\caption{An example of a manifold with a nontrivial spin structure around a homology cycle; see Coser \textit{et al}.~\cite{coser2016spin}.\label{fig:spin}}
\end{figure}

\subsection{Entanglement Negativity Measurement Protocols}

By combining the operator identities, Eqs.~(\ref{Uneg}), (\ref{mixed}), we obtain our \emph{second main result},
\begin{multline}\label{mainres2}
f(\{N\})=\delta_{(N_\mathrm{avg}^{A_1}-N_\mathrm{avg}^{A_2})\in\mathbb{Z}}\prod_{k=1}^n e^{\frac{2\pi ik}{n} \big(N_{k}^{A_1}-N_{k}^{A_2}\big)} \\
\times\begin{cases} (-1)^{N_\mathrm{avg}^{A_1}-N_\mathrm{avg}^{A_2}}~ &    n~\mathrm{even},\\
\phantom{(-}1 &   n~\mathrm{odd}.
\end{cases}
\end{multline}
The above analysis implies that measurements of \(\hat{U}_A^\mathrm{neg}\) on the FTed system yield the \Renyi entanglement negativity.

As an example, in order to measure the 3\textsuperscript{rd} \Renyi negativity one would have to average over a set of measurements using the standard FT, \(\hat{F}=\hat{F}_{+++}\), with weight \((-\frac{1}{2})\), and a set of measurements using a generalized FT, \(\hat{F}_{++-}\), with weight \((+\frac{3}{2})\); see Eq.~(\ref{neg3}).

\subsection{Spin Structures}\label{sec:spin}
The above calculation of negativities is just a special case of a general fermionic system with nontrivial spin structure.
Similar quantities appear in, e.g., string theory~\cite{SEIBERG1986272} or the classification of interacting topological systems~\cite{Shapourian2017Many,Shiozaki2018Many}. We use the connection of such manifolds to entanglement measures of fermionic systems~\cite{coser2016spin} in order to show that they are not mere theoretical constructs, but could actually be measured with ultracold atoms.
As shown in Ref.~\onlinecite{coser2016spin}, spin structures of many (1+1)-dimensional manifolds are related to polynomials in \(\hat{\rho}_{A},\hat{\rho}_A^+,\hat{P}^{\scriptscriptstyle A_\alpha}_k,\mathrm{~and~}\hat{P}^{\scriptscriptstyle B_\beta}_k\), where \(B_\beta\) are subregions of \(B\) connecting subregions \(A_\alpha\) in copy \(k=1\ldots n\).

Using the results of this paper, one may experimentally simulate spin structure partition functions on arbitrary genus manifolds, as depicted in Fig.~\ref{fig:spin}, by performing additional parity measurements, \(\hat{P}_k^{\scriptscriptstyle B_\beta}=(-1)^{\hat{N}_k^{B_\beta}}\), in region \(B_\beta\) of copy \(k\).
Similar protocols apply to ($d$+1)-dimensional manifolds obtained by cutting and gluing $\mathbb{R}^{d+1}$ at given constant-time hyperplanes.

\section{Example}\label{sec:num}
As an example simulating an experimental implementation of our protocols we study the Klich-Levitov quench model~\cite{Klich2009Quantum,klich2009many,klich2008scaling} of a tight-binding fermionic chain of length $L$ at half-filling which is initially disconnected at the middle at times ${t<0}$ and later connected at times ${t>0}$, see Fig.~\ref{fig:quench}(top). The time-dependent Hamiltonian is given by
\begin{align}
&\hat{H}(t)=\begin{cases}
\hat{H}_0 & t<0,\\
\hat{H}_1 & t>0,
\end{cases}\\
&\hat{H}_0=\sum\limits_{j=1}^{L/2-1}(\hat{a}^\dagger_j \hat{a}^{\phantom{|}}_{j+1}+\mathrm{h.c.})+\sum\limits_{j=L/2+1}^{L-1}(\hat{a}^\dagger_j \hat{a}^{\phantom{|}}_{j+1}+\mathrm{h.c.}), \nonumber\\
&\hat{H}_1=\sum\limits_{j=1}^{L-1}(\hat{a}^\dagger_j \hat{a}^{\phantom{|}}_{j+1}+\mathrm{h.c.}). \nonumber
\end{align}
The time-dependent density matrix at times ${t>0}$ both before and after the (generalized) Fourier transform are thus given by
\begin{eqnarray}
\hat{\rho}(t)&=&\lim\limits_{\beta\to\infty}e^{-i\hat{H}_1t}\frac{e^{-\beta\hat{H}_0}}{\Tr\, e^{-\beta\hat{H}_0}}e^{i\hat{H}_1t},\\
\widetilde{\hat{\rho}^\otn}(t)&=&\hat{F}_{\{\sigma\}}^{\phantom{|}}\hat{\rho}(t)^\otn\hat{F}^\dagger_{\{\sigma\}}.
\end{eqnarray}

In order to simulate our measurement protocols, we must determine the probability $P[f(\{N\})]$ of a result for $f(\{N\})$ in our measurement protocols, Eqs.~(\ref{mainres}), (\ref{mainres2}).
\begin{align}
& P\big[f(\{N\})=e^{\frac{2\pi i m}{n}}\big]=\Tr\Big\{\delta_{f(\{\hat{N}\}),\exp(\frac{2\pi i m}{n})}\widetilde{\hat{\rho}^\otn}\Big\} \nonumber\\
&={\sum\limits_{q,q'=1}^n}  e^{-\frac{2\pi i m q}{n}}\,\Tr\Big\{{\prod\limits_{k=1}^{n}} e^{\frac{2\pi i}{n} \big(\hat{N}_{k}^{A_1}\pm \hat{N}_{k}^{A_2}\big)[[k(+\frac{1}{2})]q+q']}\widetilde{\hat{\rho}^\otn}\Big\}.
\end{align}
Here, the ``$\pm$" distinguishes between the negativity and the entropy, and the ``$(+\frac{1}{2})$" applies only for even $n$.

The evaluation of this trace of exponentials is done using the methods of Refs.~\onlinecite{Klich2002FCS,Klich2014FCS,GoldsteinSela2018} for the entanglement entropy, and using the methods of Refs.~\onlinecite{eisler2015partial,Eisler2016Entanglement,cornfeld2018imbalance} for the entanglement negativity.
We simulate \(A\cup B\) of 32 sites, with ${|A|=16}$ for the entropy and ${|A_{1,2}|=2}$ for the negativity.
The data presented in Fig.~\ref{fig:quench}(bottom) is attained by randomly sampling $f(\{N\})$ from $P[f(\{N\})]$ and averaging over the instances. 
We sample 600 instances for the \Renyi entanglement entropy as well as for the \Renyi entanglement negativity. 
The error bars display the statistical standard error of the mean (SEM).



\section{Conclusions}\label{sec:conc}
We described, for the first time, experimental protocols for direct detection of various entanglement measures in fermionic systems beyond the 2\textsuperscript{nd} \Renyi entropy. 
We presented a complete paradigm that enables the evaluation of all \Renyi entropies and negativities for both pure and mixed states.

Possible applications include: testing the scaling laws of entanglement in many-body quantum states, determining the coherence of quantum simulators, and incorporating entanglement detections in quantum computations. Our protocols also allows the use of cold atoms to quantum simulate the partition function of fermions on certain manifolds with nontrivial spin structures.

All our protocols are based on performing Fourier transforms on the quantum system, which may be realized by applying sequences of beam splitters, and are readily attainable using existing technologies. 

\section*{Acknowledgements}  	  
E.S. was supported in part by the Israel Science Foundation (Grant No. 1243/13) and by the US-Israel Binational Science Foundation (Grant No. 2016255). M.G. was supported by the Israel Science Foundation (Grant No. 227/15), the German Israeli Foundation (Grant No. I-1259-303.10), the US-Israel Binational Science Foundation (Grant No. 2016224), and the Israel Ministry of Science and Technology (Contract No. 3-12419).

\appendix
\renewcommand{\theequation}{A\arabic{equation}}
\setcounter{equation}{0}
\section{Pure Monomials}\label{app:pure}
We herein explicitly derive Eq.~(\ref{Uneg}).

We wish to evaluate the action of shift operator over a pure monomial of the decomposed reduced density matrix \({\Tr\{(\hat{\rho}_A^{\pm})^n\} = \Tr\{\hat{V}_A(\hat{\rho}_A^{\pm})^\otn\}}\); see Eq.~(\ref{T2def}). We thus expand it in the local Majorana basis,
\begin{gather}
(\hat{\rho}_A^{\pm})^\otn=\sum\limits_{\{\bm{\mu}\}}w_{\{\bm{\mu}\}}(\pm i)^{\sum_k|\bm{\mu}^2_k|}\underset{k}{\overrightarrow{\prod}}(\hat{\bm{\gamma}}^{\sA{1}}_k)^{\bm{\mu}^{1}_k}(\hat{\bm{\gamma}}^{\sA{2}}_k)^{\bm{\mu}^{2}_k},\\
w_{\{\bm{\mu}\}}\eqdef\prod\limits_k w_{\bm{\mu}^{1}_k,\bm{\mu}^{2}_k},
\end{gather} 
using the directional product, \({\vec{\prod}_k \hat{a}_k=\hat{a}_1\hat{a}_2\cdots\hat{a}_n}\).
Let us look at the tensor product after a Fourier transform,
\begin{equation}
\hat{F}\hat{\rho}_A^\otn\hat{F}^\dagger \eqdef\sum\limits_{\{\bm{\mu}\}}\tilde{w}_{\{\bm{\mu}\}}\underset{k}{\overrightarrow{\prod}}(\hat{\bm{\gamma}}^{\sA{1}}_k)^{\bm{\mu}^{1}_k}(\hat{\bm{\gamma}}^{\sA{2}}_k)^{\bm{\mu}^{2}_k}.
\end{equation}
Since the total number of Majorana operators in any site is invariant under the transformation, as shown in Sec.~\ref{sec:gauss}, one finds  that the phase factor remains unchanged, \textit{i.e.},
\begin{equation}
\hat{F} (\hat{\rho}_A^{\pm})^\otn\hat{F}^\dagger=\sum\limits_{\{\bm{\mu}\}}\tilde{w}_{\{\bm{\mu}\}}(\pm i)^{\sum_k|\bm{\mu}^2_k|}\underset{k}{\overrightarrow{\prod}}(\hat{\bm{\gamma}}^{\sA{1}}_k)^{\bm{\mu}^{1}_k}(\hat{\bm{\gamma}}^{\sA{2}}_k)^{\bm{\mu}^{2}_k}.
\end{equation}
We are now in a position to look at the action of
\begin{equation}
\hat{U}_{A_{1,2}}^\mathrm{fer}=\prod_{k=-(n-1)/2}^{(n-1)/2} \omega^{k\hat{N}_{k}^{A_{1,2}}}.
\end{equation}
It can be expressed in terms of Majorana operators,
\begin{equation}
\hat{\mathbf{a}}_k^{\sA{1,2}}=\tfrac{1}{2}(\hat{\bm{\gamma}}_k^{\sA{1,2}}+i\hat{\bm{\gamma}}'^{\sA{1,2}}_k),
\end{equation}
such that,
\begin{equation}
\begin{aligned}
(\hat{U}_{A_\alpha}^\mathrm{fer})^{\pm 1}&=\prod\limits_{j\in A_\alpha}\prod\limits_k\big(\tfrac{\omega^{\pm k}+1}{2}+\tfrac{\omega^{\pm k}-1}{2}i[\hat{\gamma}^{\sA{\alpha}}_k]_j[\hat{\gamma}'^{\sA{\alpha}}_k]_j\big) \\
&=\prod\limits_{j\in A_\alpha}\prod\limits_k\tfrac{\omega^{\pm k}+1}{2}\big(1+\tfrac{\omega^{\pm k}-1}{\omega^{\pm k}+1}i[\hat{\gamma}^{\sA{\alpha}}_k]_j[\hat{\gamma}'^{\sA{\alpha}}_k]_j\big) \\
&=\prod\limits_{j\in A_\alpha}\tfrac{1}{2^{n-1}}\prod\limits_k\big(1\mp\tan(\tfrac{\pi k}{n})[\hat{\gamma}^{\sA{\alpha}}_k]_j[\hat{\gamma}'^{\sA{\alpha}}_k]_j\big).
\end{aligned}
\end{equation}
We observe two main properties of this expression: (i) the difference between \(\hat{U}_{A_2}^\mathrm{fer}\) and \((\hat{U}_{A_2}^\mathrm{fer})^{-1}\) is just a minus sign for any pair of Majorana operators in region \(A_2\); (ii) within the trace \({\Tr\{\sum_{\bm{\mu}}w_{\bm{\mu}}(\hat{\bm{\gamma}})^{\bm{\mu}}\}=w_0}\), and hence only terms with even number of Majorana operators contribute. Using the these properties we find
\begin{align}
& \Tr\{(\hat{\rho}_{A}^\pm)^n\}=\textstyle \Tr\{\hat{V}_A(\hat{\rho}_A^{\pm})^\otn\}=\Tr\{\hat{U}_A^\mathrm{fer}\hat{F}(\hat{\rho}_A^\pm)^\otn\hat{F}^\dagger\} \nonumber\\
&=\Tr\sum\limits_{\{\bm{\mu}\}}\tilde{w}_{\{\bm{\mu}\}}(\pm i)^{\sum_k|\bm{\mu}^2_k|}\hat{U}_A^\mathrm{fer}\underset{k}{\overrightarrow{\prod}}(\hat{\bm{\gamma}}^{\sA{1}}_k)^{\bm{\mu}^{1}_k}(\hat{\bm{\gamma}}^{\sA{2}}_k)^{\bm{\mu}^{2}_k} \nonumber\\
&=\Tr\!\sum\limits_{|\bm{\mu}^{1,2}_k|\ \mathrm{even}}\!\!\tilde{w}_{\{\bm{\mu}\}}(-1)^{\frac{1}{2}\sum_k|\bm{\mu}^2_k|}\hat{U}_A^\mathrm{fer}\underset{k}{\overrightarrow{\prod}}(\hat{\bm{\gamma}}^{\sA{1}}_k)^{\bm{\mu}^{1}_k}(\hat{\bm{\gamma}}^{\sA{2}}_k)^{\bm{\mu}^{2}_k} \nonumber\\
&=\Tr\!\sum\limits_{|\bm{\mu}^{1,2}_k|\ \mathrm{even}}\!\!\tilde{w}_{\{\bm{\mu}\}}\hat{U}_{A_1}^\mathrm{fer}(\hat{U}_{A_2}^\mathrm{fer})^{-1}\underset{k}{\overrightarrow{\prod}}(\hat{\bm{\gamma}}^{\sA{1}}_k)^{\bm{\mu}^{1}_k}(\hat{\bm{\gamma}}^{\sA{2}}_k)^{\bm{\mu}^{2}_k} \nonumber\\
&=\Tr\{\hat{U}_{A_1}^\mathrm{fer}(\hat{U}_{A_2}^\mathrm{fer})^{-1}\hat{F}\hat{\rho}^\otn\hat{F}^\dagger\}.
\end{align}
Specifically,
\begin{equation}
\Tr\{(\hat{\rho}_{A}^\pm)^n\} = \Tr\{\hat{U}_{A_1}^\mathrm{fer}(\hat{U}_{A_2}^\mathrm{fer})^{-1}(\hat{F}\hat{\rho}^\otn\hat{F}^\dagger)\}.
\end{equation}

\subsection{Gaussian Transformations}\label{sec:gauss}
We herein show a useful property of any Gaussian transformation such as (generalized) Fourier transforms.

Any Gaussian transformation, \({\hat{G}\propto e^{-\sum_{\mu\nu}\hat{\gamma}_\mu \mathrm{M}_{\mu\nu}\hat{\gamma}_\nu}}\), acts on a single Majorana by
\begin{equation}
\hat{G}\hat{\gamma}_\mu\hat{G}^\dagger=\sum_{\nu}[\Omega_G]_{\mu\nu}\hat{\gamma}_\nu.
\end{equation}
One of its properties is that when acting on some monomial of degree \(q\) of (distinct) Majorana operators such as \({\hat{\Gamma}_{\{\mu\}_q} = \hat{\gamma}_{\mu_1}\hat{\gamma}_{\mu_2}\cdots\hat{\gamma}_{\mu_q}}\), 
it preserves the degree.

Proof by contradiction: Assume that there is some monomial \({\hat{\Gamma}_{\{\nu\}_p} = \textstyle\hat{\gamma}_{\nu_1}\hat{\gamma}_{\nu_2}\cdots\hat{\gamma}_{\nu_p}}\)of degree \(p<q\) that would appear in the transform. This would imply that \({\Tr\{\hat{\Gamma}_{\{\nu\}_p}\hat{G}\hat{\Gamma}_{\{\mu\}_q}\hat{G}^\dagger\}\neq 0}\), however, one has
\begin{multline}
\Tr\{\hat{\Gamma}_{\{\nu\}_p}\hat{G}\hat{\Gamma}_{\{\mu\}_q}\hat{G}^\dagger\} = \Tr\{\hat{G}^\dagger\hat{\Gamma}_{\{\nu\}_p}\hat{G}\hat{\Gamma}_{\{\mu\}_q}\}\\
=\sum\limits_{\{\nu'\}_{p'} \ : \ p'\le p} \tilde{w}_{\{\nu'\}}\Tr\{\hat{\Gamma}_{\{\nu'\}_{p'}}\hat{\Gamma}_{\{\mu\}_q}\}\underset{p'\neq q}{=}0.
\end{multline}
{\flushright
Q.E.A.\par}
We thus conclude that
\begin{equation}
\hat{G}\hat{\Gamma}_{|\{\mu\}|=q}\hat{G}^\dagger = \sum\limits_{|\{\mu'\}|=q} \tilde{w}_{\{\mu'\}}\hat{\Gamma}_{\{\mu'\}}.
\end{equation}


%

\end{document}